\documentclass{elsart}
\usepackage{epsfig}

\begin{document}

\begin{frontmatter}

\title{\Large\bf Measurements of the Mass and Full-Width of the
                 {\boldmath $\eta_c$} Meson}

\date{6 Jan. 2003}

\maketitle

\begin{center}

J.~Z.~Bai$^1$,        Y.~Ban$^{9}$,          J.~G.~Bian$^1$,
X.~Cai$^{1}$,          J.~F.~Chang$^1$,
H.~F.~Chen$^{16}$,    H.~S.~Chen$^1$,
Jie~Chen$^{8}$,        J.~C.~Chen$^1$,     
Y.~B.~Chen$^1$
% ...
\end{center}
\vskip 0.3cm
\begin{center}
(The BES Collaboration)
\end{center}
\vskip 0.3cm

\small
\begin{center}
$^1$ Institute of High Energy Physics, Beijing 100039, People's Republic of
     China\\
% ...
$^8$ Nankai University, Tianjin 300071, People's Republic of China\\
$^{9}$ Peking University, Beijing 100871, People's Republic of China\\
% ...
$^{16}$ University of Science and Technology of China, Hefei 230026,
        People's Republic of China\\
% ...

\vspace{0.2cm}

\end{center}

\normalsize

\begin{abstract}
In a sample of 58 million $J/\psi$ events collected with the
BES II detector, the process J/$\psi\to\gamma\eta_c$ is
observed in five different decay channels:
$\gamma K^+K^-\pi^+\pi^-$, $\gamma\pi^+\pi^-\pi^+\pi^-$,
$\gamma K^\pm K^0_S \pi^\mp$ (with $K^0_S\to\pi^+\pi^-$),
$\gamma \phi\phi$ (with $\phi\to K^+K^-$) and $\gamma p\bar{p}$. 
From a combined fit of all five channels,
we determine the mass and full-width of $\eta_c$ to be
$m_{\eta_c}=2977.5\pm1.0~(\mbox{stat.})\pm1.2~(\mbox{syst.})$ MeV/$c^2$
and $\Gamma_{\eta_c} = 17.0\pm3.7~(\mbox{stat.})\pm7.4~(\mbox{syst.})$
MeV/$c^2$.

\vspace{3\parskip}

\end{abstract}

\end{frontmatter}
\clearpage

Since 1980, numerous efforts have been made to determine the mass and
full width of the $\eta_c$ [1--11].  However, from a theoretical point
of view, the accuracies of these experimental measurements are still
not sufficient. For instance, in order to calculate the strength of
the spin-spin interaction term in non-relativistic potential models,
it is necessary to know precisely the mass difference between the
J/$\psi(1^{--})$ and $\eta_c(0^{-+})$.  While the mass of the $J/\psi$ is
determined with high accuracy to be 3096.88$\pm$0.04 MeV/$c^2$, the
$\eta_c$ mass is measured with much less accuracy to be $ 2979.7 \pm
1.5$~MeV/$c^2$, an average by the Particle Data Group (PDG)~\cite{pdg2002}
of 10 measurements with an internal confidence level of only 0.001.  Different
measurements of the full width of the $\eta_c$ also have poor internal
consistency.  The PDG \cite{pdg2002} determines an
average value for the $\eta_c$ full width of $16.0^{+3.6}_{-3.2}$
MeV/$c^2$ from six experiments, whose experimental results vary from 7
MeV/$c^2$ to 27 MeV/$c^2$, with large errors.  Such an accuracy is
inadequate for some studies of charmonium physics~\cite{charm}
and additional, more precise measurements of both $m_{\eta_c}$ and
$\Gamma_{\eta_c}$ are needed.

The $\eta_c$ mass and width have been measured previously by the 
BES collaboration with data samples of
3.79 million $\psi$(2S) events \cite{bespsi} and 7.8 million
$J/\psi$ events \cite{besjpsi} collected with the BES I detector \cite{bes1}. 
In the latter sample, the process J/$\psi\to\gamma\eta_c$ was
observed in five different $\eta_c$  decay channels:
$K^+K^-\pi^+\pi^-$, $\pi^+\pi^-\pi^+\pi^-$,
$K^\pm K^0_S \pi^\mp$ (with $K^0_S\to\pi^+\pi^-$),
$\phi\phi$ (with $\phi\to K^+K^-$) and
$K^+K^-\pi^0$, and
the mass of the $\eta_c$ was determined to be
$2976.6\pm2.9~(\mbox{stat.})\pm1.3~(\mbox{syst.})$ MeV/$c^2$. 
%(Remove this sentence starting from "In the later sample..."  ??)
Combined with the results from $\psi(2S)\to\gamma\eta_c$,
the mass and the full width of $\eta_c$ were determined to be
$m_{\eta_c} = 2976.3\pm2.3~(\mbox{stat.})\pm1.2~(\mbox{syst.})$ MeV/$c^2$
and $\Gamma_{\eta_c} = 11.0\pm 8.1~(\mbox{stat.})\pm 4.1~(\mbox{syst.})$
MeV/$c^2$.

In this paper we present results with much higher statistics using
a recent sample of 58 million $J/\psi$ events obtained with the upgraded
BESII detector \cite{bes2}.
The upgrade from BES I to BES II includes the replacement of
the inner drift chamber with a straw-tube vertex chamber (VC), composed of
12 tracking layers arranged around a beryllium beam pipe and
with a spatial resolution of about 90 $\mu$m;
a new barrel time-of-flight counter (BTOF) with a
time resolution of 180 ps; and a new main
drift chamber (MDC), which has 10 tracking layers providing
a $dE/dx$ resolution of $\sigma_{dE/dx} = 8.4\%$ 
and a momentum resolution of $\sigma_p/p = 1.7\% \sqrt{1+p^2}$ ($p$ in GeV) 
for charged tracks. These upgrades augment the pre-existing calorimeter and
muon tracking systems.  The barrel shower counter
(BSC), which covers $80\%$ of $4\pi$ solid angle, has an energy resolution
of $\sigma_E/E = 22\%/\sqrt{E}$ ($E$ in GeV) and a spatial resolution
of 7.9 mrad in $\phi$ and 2.3 cm in $z$.
The $\mu$ identification system consists of three
double layers of proportional tubes interspersed in the iron flux return
of the magnet. They provide coordinate measurements along the muon
trajectories with resolutions of 3 cm and 5.5 cm in $\phi$ and $z$,
respectively.  

The $\eta_c$ mass and width are measured using the reactions
J/$\psi\to\gamma\eta_c$; $\eta_c\to K^+K^-\pi^+\pi^-$, $\pi^+\pi^-\pi^+\pi^-$,
$K^\pm K^0_S \pi^\mp$ (with $K^0_S\to\pi^+\pi^-$), 
$\phi\phi$ (with $\phi\to K^+K^-$) and $p\bar{p}$.
Event selection criteria for each channel are described in detail in
our previous papers \cite{dly2kp,dly2pp,dlyk3p}.
Here we repeat only the essential information and emphasize those
considerations that are unique to the $m_{\eta_c}$ and $\Gamma_{\eta_c}$
measurements.

Candidate events are required to have the correct number of
charged tracks for a given hypothesis. 
Each
track must be well fit to a helix in the polar angle
range $|\cos\theta| < 0.84$ and have a transverse momentum
above $60$ MeV/c.
For the decay channels 
$J/\psi \to \gamma K^+K^-\pi^+\pi^-$,
    $J/\psi \to \gamma \pi^+\pi^-\pi^+\pi^-$,
    $J/\psi \to \gamma K^\pm\pi^\mp\pi^+\pi^-$ and
    $J/\psi \to \gamma p \bar{p}$,
at least one photon with energy $E_{\gamma}> 30$~MeV is required in the
barrel shower counter 

Events are kinematically fitted with four constraints (4C) to the hypotheses:
$J/\psi \to \gamma K^+K^-\pi^+\pi^-$,
$J/\psi \to \gamma \pi^+\pi^-\pi^+\pi^-$,
$J/\psi \to \gamma K^\pm\pi^\mp\pi^+\pi^-$, and
$J/\psi \to \gamma p \bar{p}$.
A one-constraint(1C) fit is performed for the
$J/\psi \to \gamma_{miss} K^+K^-K^+K^-$ hypothesis,
where $\gamma_{miss}$ indicates that this photon is not detected.
Events with a $\chi^2$ less than 40.0 for a particular channel are selected.

In order to remove backgrounds from non-radiative decay channels,
all selected events are subjected to a kinematic fit
with four constraints to the hypotheses:
$J/\psi \to K^+K^-\pi^+\pi^-$,
$J/\psi \to \pi^+\pi^-\pi^+\pi^-$ and
$J/\psi \to K^\pm\pi^\mp\pi^+\pi^-$.
Backgrounds from the $J/\psi$ peak are 
removed by requiring that
$\chi^2(J/\psi \to K^+K^-\pi^+\pi^-)>20.0$ (for $K^+K^-\pi^+\pi^-$);
$\chi^2(J/\psi \to \pi^+\pi^-\pi^+\pi^-)>10.0$ (for
$\pi^+\pi^-\pi^+\pi^-$) and
$\chi^2(J/\psi \to K^\pm\pi^\mp\pi^+\pi^-)>10.0$ (for $K^\pm
K_{S}^{0}\pi^\mp$).
For the $J/\psi \to \gamma p \bar{p}$ channel,
we require that the opening angle of the two charged tracks is smaller than
$179^\circ$.
A detailed Monte Carlo simulation shows that these cuts, referred to below as the
$J/\psi$ veto, do not distort 
the invariant mass distributions around the $\eta_c$ signal peak.

Two additional variables are used to reject events with wrong
final state assignments. The first variable,
$|U_{miss}|=|E_{miss}-P_{miss}|$, is used to reject events with
multi-photons and misidentified charged particles. Here,
$E_{miss}$ and $P_{miss}$ are, respectively, the missing energy
and momentum calculated using measured quantities for
charged tracks. A second variable,
$P_{t\gamma}^2 = 4|P_{miss}|^2\sin^2(\theta_{t\gamma}/2)$,
where $\theta_{t\gamma}$ is the angle between the missing momentum
and the photon direction,
is used to reduce backgrounds from $\pi^0$'s.
The specific values of the selection requirements for these two
kinematic variables are summarized in Table~\ref{two-cuts}.
Additional requirements to remove backgrounds from a few specific channels
are summarized in Table~\ref{cut-bg}.

\begin{table}[h]
\caption{Cuts imposed on $|U_{miss}|$ and $P_{t\gamma}^2$
         for event selection.}
\label{two-cuts}
\begin{tabular}{|c|c|c|}
\hline
mode ($J/\psi \rightarrow \gamma X$ ) & $\mid U_{miss} \mid$ (GeV/$c^2$) 
& $P_{t\gamma}^2$ [$(\mbox{GeV}/c)^2$) \\ \hline
$ \gamma K^+K^-\pi^+\pi^- $      & $<0.15$  &  $<0.002$  \\ \hline
$ \gamma \pi^+\pi^-\pi^+\pi^- $  & $<0.10$  &  $<0.0015$ \\ \hline
$ \gamma K^\pm K_{S}^{0}\pi^\mp$ ($\gamma K^\pm\pi^\mp\pi^+\pi^-$)  & --  &  $<0.003$ \\ \hline
$ \gamma p \bar{p}$  & $<0.15$  &  $<0.003$ \\ \hline
\end{tabular}
\end{table}

\begin{table}[htb]
\caption{Cuts to remove backgrounds from specific channels.}
\label{cut-bg}
\begin{tabular}{|c|c|c|}
\hline
mode ($J/\psi \rightarrow \gamma X$ ) & cut &  background\\ \hline
$ \gamma K^+K^-\pi^+\pi^- $      & 
   $|M_{\pi^+\pi^-\pi^0}-M_\omega|>40$ MeV/$c^2$ & $J/\psi\to \omega K^+K^-$     \\ \hline
$ \gamma K^+K^-\pi^+\pi^- $      & 
   $|M_{K^+K^-}-M_\phi|>20$ MeV/$c^2$ & $J/\psi\to \phi \pi^+\pi^-$     \\ \hline
$ \gamma \pi^+\pi^-\pi^+\pi^- $  &
   $|M_{\pi^+\pi^-\pi^0}-M_\omega|>40$ MeV/$c^2$ & $J/\psi\to \omega \pi^+\pi^-$ \\ \hline
% & on both $M_{\pi^+\pi^-}$, & \\
$ \gamma \pi^+\pi^-\pi^+\pi^- $  &
 $|M_{\pi^+\pi^-}-M_{K_S^0}|>25$ MeV/$c^2$ & $J/\psi\to \gamma K_S^0 K_S^0$ \\ \hline
\end{tabular}
\end{table}

For $J/\psi \to \gamma \pi^+\pi^-\pi^+\pi^-$ candidate events
with more than one $\gamma$, we suppress $\pi^0$ background by requiring
that $\mid M(\gamma_1 \gamma_2) - M(\pi^0) \mid > 60 $ MeV/$c^2$
if $\vec{P}_{miss}$ is in the same plane as the
two photons $\gamma_1$ and $\gamma_2$,
i.e. $\hat{P_{miss}} \cdot (\hat{r_{\gamma1}} \times \hat{r_{\gamma2}}) <0.15$. 
Here, $\hat{P_{miss}}$ is the unit vector of the missing momentum for all
charged tracks; $\hat{r_{\gamma1}}$ and $\hat{r_{\gamma2}}$ are unit
vectors for the $\gamma_1$ and $\gamma_2$ directions determined from the BSC;
and $M(\gamma_1 \gamma_2)$ is the invariant mass of $\gamma_1 \gamma_2$.
When we calculate $M(\gamma_1 \gamma_2)$, 
it is assumed that the missing particle decays to $\gamma_1$ and $\gamma_2$,
and $M(\gamma_1 \gamma_2)$ can be obtained by using
$P_{miss}$ and the angles between $\vec{P}_{miss}$ and the $\gamma$ direction.
The advantage of this technique is that it uses the momenta 
of charged tracks measured by the MDC, which has good momentum resolution, 
and is independent of the photon energy measurement.
For $J/\psi\to \gamma K^+K^-\pi^+\pi^-$, $\gamma K^\pm K^0_S \pi^\mp$, and
$\gamma p\bar{p}$,
we require that $\mid M(\gamma_1 \gamma_2) - M(\pi^0) \mid ~> 50 $ MeV/$c^2$
when $\hat{P_{miss}} \cdot (\hat{r_{\gamma1}} \times \hat{r_{\gamma2}}) <0.14$.

For the $K^\pm K_{S}^{0}\pi^\mp$ (with $K^0_S\to\pi^+\pi^-$) channel, the
$\pi^+\pi^-$ invariant mass for the
$K^0_S $ candidate is required to be within 25~MeV/$c^2$
of the $K_{S}^{0}$ mass.
For the $\phi\phi$ (with $\phi\to K^+K^-$) channel,
the invariant masses of both candidate $\phi$'s, corresponding to
$K^+K^-$ pairs, are required to be within 20~MeV/$c^2$ of the $\phi$ mass.

\begin{figure}[htbp]
  \centering
  \begin{minipage}[b]{0.48\linewidth}
    \centering
    \includegraphics[width=6.5cm]{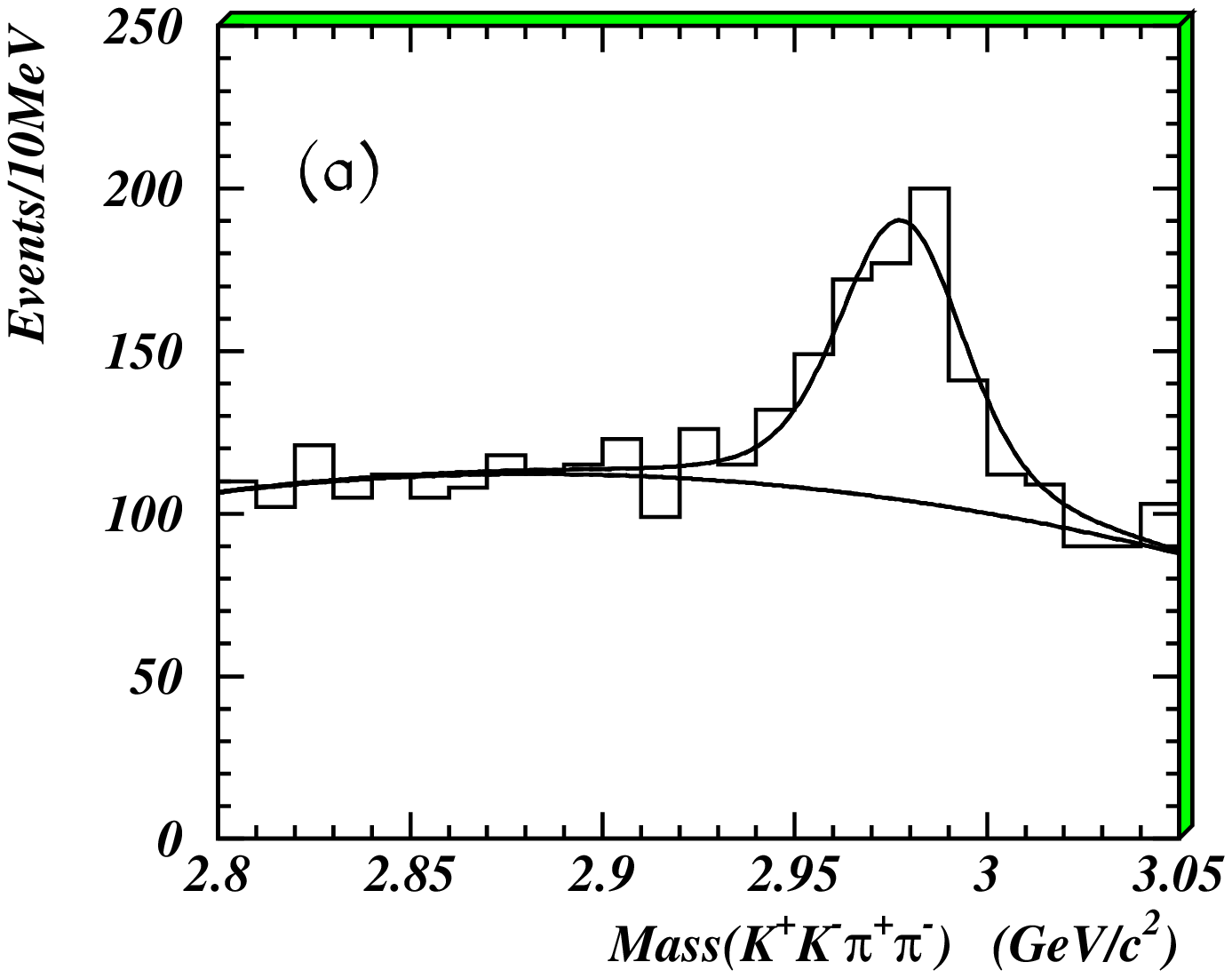}
  \end{minipage}
  \hfill
  \begin{minipage}[b]{0.48\linewidth}
    \centering
    \includegraphics[width=6.5cm]{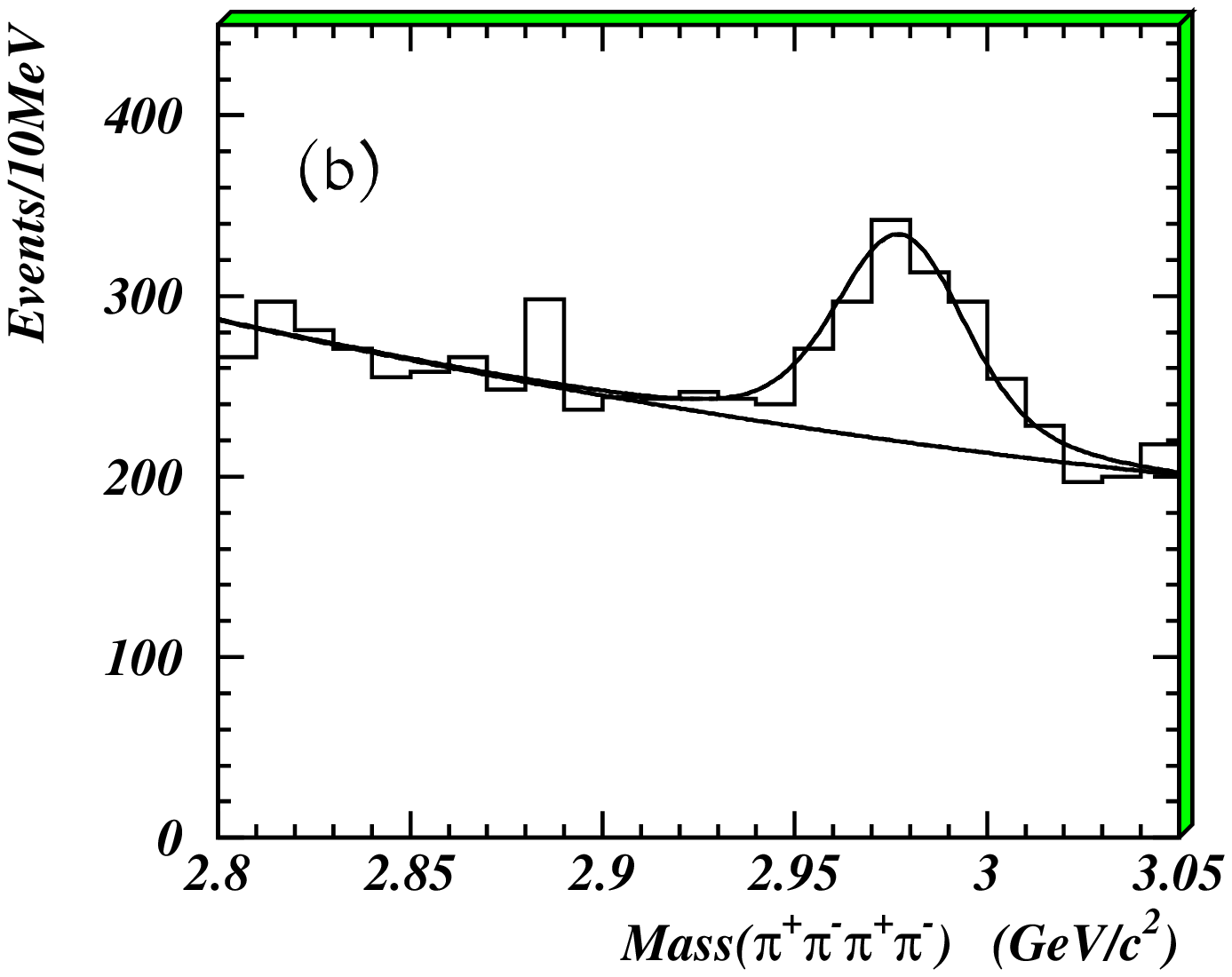}
  \end{minipage}

  \begin{minipage}[b]{0.48\linewidth}
    \centering
    \includegraphics[width=6.5cm]{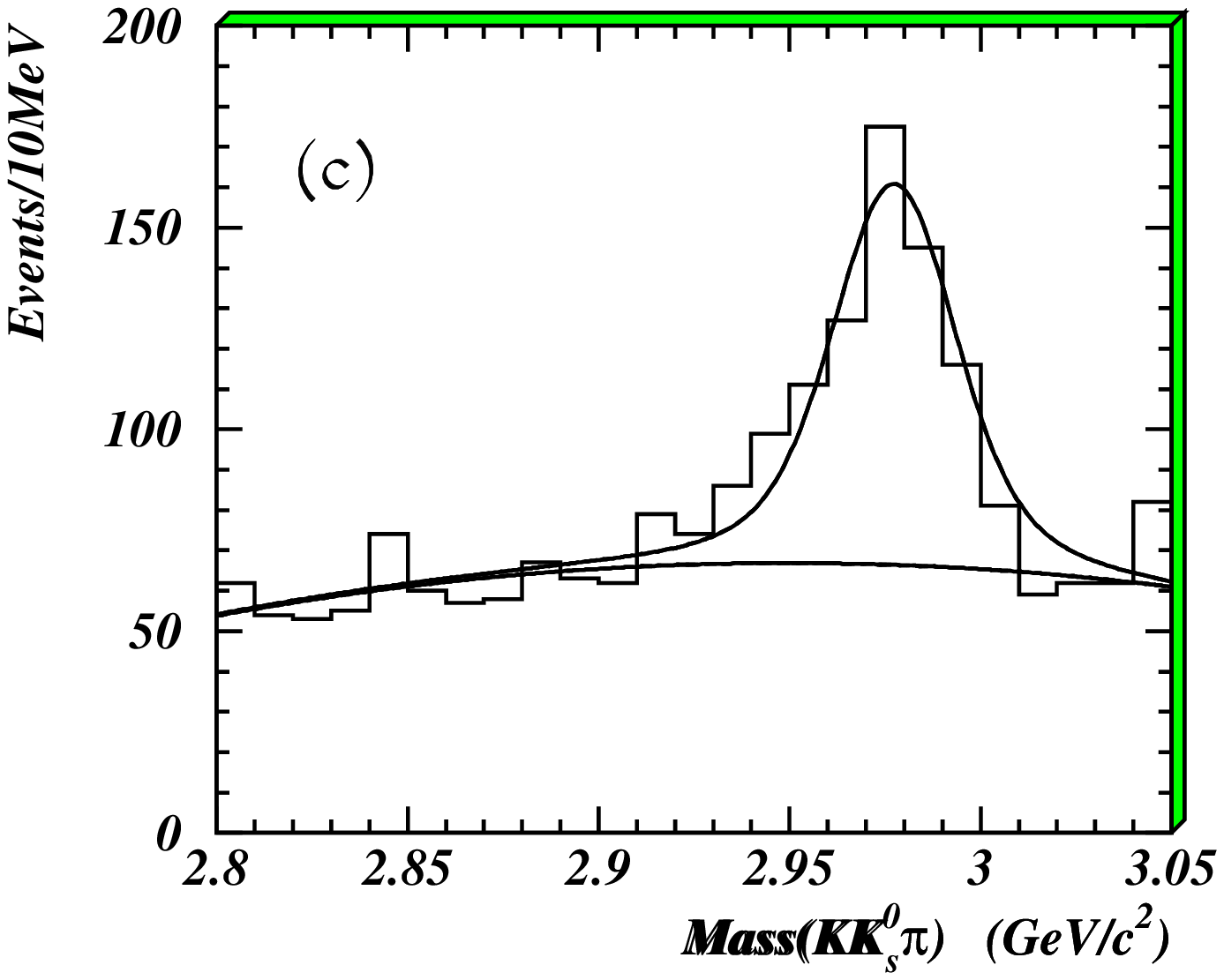}
  \end{minipage}
  \hfill
  \begin{minipage}[b]{0.48\linewidth}
    \centering
    \includegraphics[width=6.5cm]{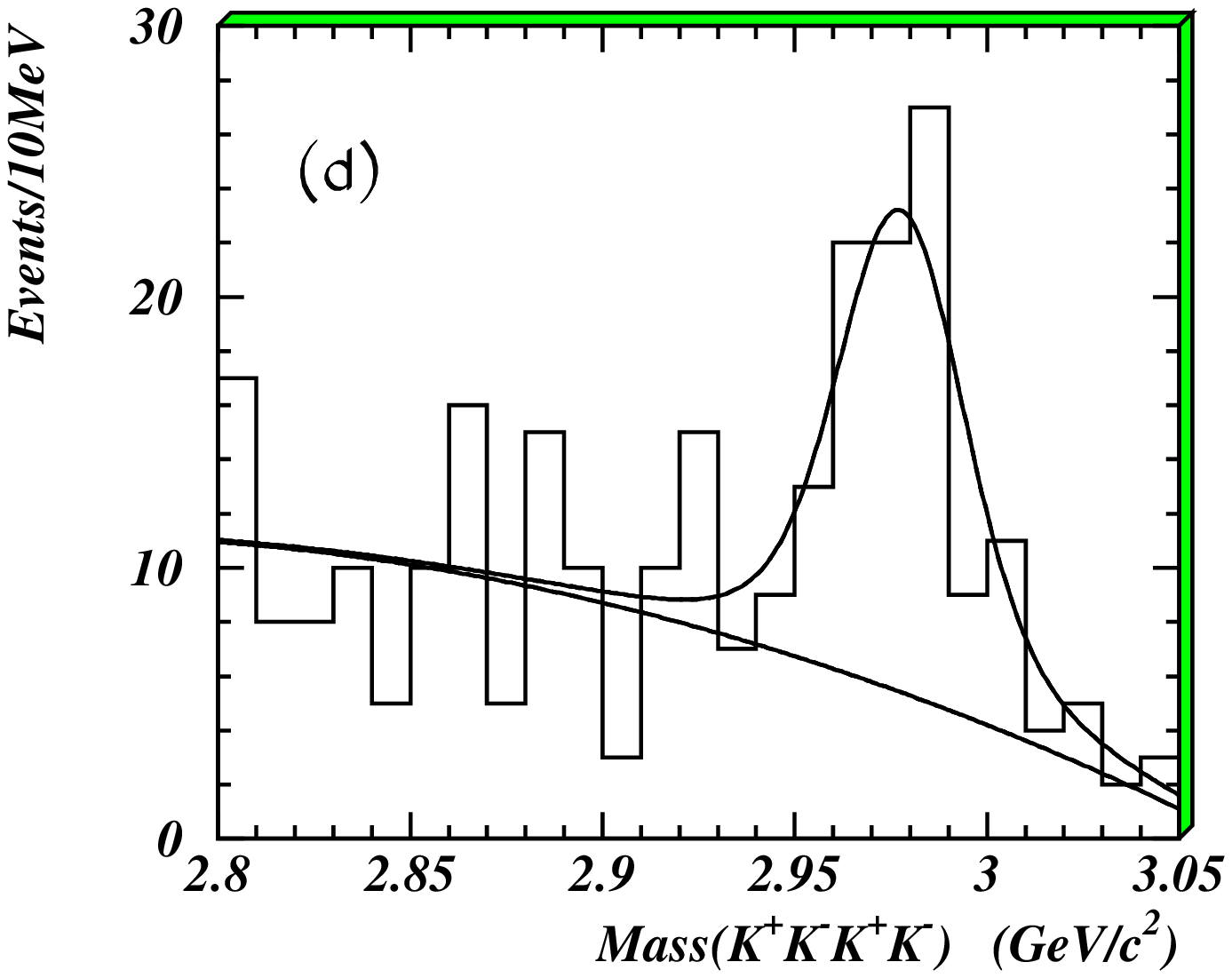}
  \end{minipage}
  \begin{minipage}[b]{0.48\linewidth}
    \centering
    \includegraphics[width=6.5cm]{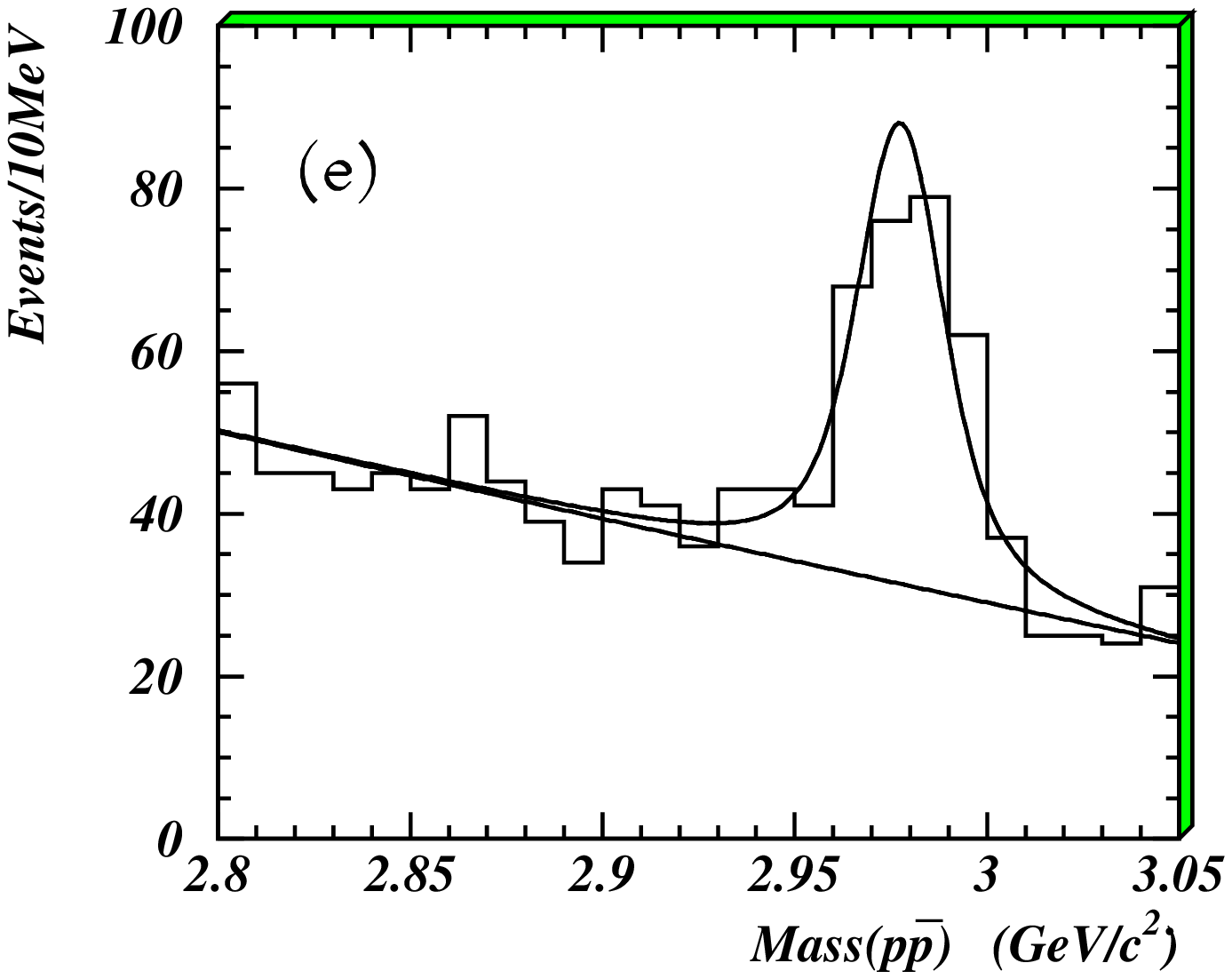}
  \end{minipage}
    \caption[]{The invariant mass distributions in the $\eta_c$ region for channels
               (a) $m_{K^+K^-\pi^+\pi^-}$, (b) $m_{\pi^+\pi^-\pi^+\pi^-}$,
                 (c) $m_{K^\pm K^0_S\pi^\mp}$, (d) $m_{\phi\phi}$ and
                 (e) $m_{p\bar{p}}$. }
    \label{fit-2xx-final}
\end{figure}

After the event selection, the invariant mass spectra
for the individual  decay modes are obtained, as shown in Fig.~\ref{fit-2xx-final}. 
An unbinned maximum likelihood fit using MINUIT~\cite{minuit} is performed
for all five channels simultaneously, with the fitting function for 
a given channel $i$ given by
$$
f_i(m) = a_i [BW(M,\Gamma,m)\otimes GS(m,\sigma_i)]\times EFF_i(m)
       + (1-a_i)BG_i(m),
$$
where $M$ and $\Gamma$ are the mass and width of the $\eta_c$, respectively,
$\sigma_i$ is the mass resolution in the $\eta_c$ region,
$BW$ is a Breit-Wigner function describing the $\eta_c$ signal,
$EFF_i$ is an efficiency correction function, and
$BG_i$ is a second-order polynomial function describing the background shape. 
In order to include the experimental resolution,
the $BW$ function is folded with a Gaussian
resolution function $GS$ with
the resolution $\sigma_i$ fixed at a value determined from the
Monte Carlo simulation.
The parameters $M$ and $\Gamma$ and the coefficients of 
the polynomial function, $a_i$, are determined from the fit.
The log likelihood function for the channel $i$ is given by
$$
S_i=-\ln{L_i} = -\ln({\displaystyle \prod_{j=1}^{N^{\rm event}_i}f_i(m_j)}),
$$
where $N^{\rm event}_i$ is the total number of events.
The overall log likelihood function, 
$$
S = \sum_{i=1}^{5} S_i, 
$$
is minimized to obtain the fitting results from the five channels 
simultaneously. 
The fit result is shown in Fig.~\ref{fit-2xx-final},
and the fitted $\eta_c$ mass and width are determined to be
$m_{\eta c} = 2977.5\pm 1.0 $ MeV/$c^2$ and
$\Gamma_{\eta c} = 17.0 \pm 3.7$ MeV/$c^2$.
The background in Fig.~\ref{fit-2xx-final}(b),(d) and (e)
can also be fitted with a linear polynomial function, 
with results that are almost the same. 

Systematic errors in determining the
$\eta_c$ mass and width originate mainly from the mass-scale calibration,
background shape, fitting range, difference between data and Monte Carlo
simulation, $J/\psi$ veto, and uncertainties associated with the mass
resolution. We use 1.5 million $\psi(2S)$ data collected during the $J/\psi$ run
to check the mass-scale calibration.
The measured $\chi_{c2}$ mass is $3555.2\pm1.4$ MeV/$c^2$
from decays $\psi(2S)\to\gamma \pi^+\pi^-\pi^+\pi^-$ and
$3560.2\pm6.0$ MeV/$c^2$ from decays $\psi(2S)\to\gamma K^+K^-\pi^+\pi^-$,
respectively. The combined weighted average is $m_{\chi_{c2}}=3555.5\pm1.4$
MeV/$c^2$,  a difference of $0.7\pm1.4$ MeV/$c^2$ from the world average
obtained by the PDG~\cite{pdg2002}. In addition, we measured
masses of the $K_{s}^{0}$, $\phi$ and $\Lambda$ from the 58 million $J/\psi$ data sample
to check the mass-scale calibration. Results of the masses and
mass differences with PDG values~\cite{pdg2002} are given
in Table~\ref{Fit-inc}. The systematic error on the 
overall mass scale is estimated to be 0.8 MeV/$c^2$.

\begin{table}[htb]
\caption{Comparison of $K_s^0$, $\phi$ and $\Lambda$ mass peak positions.}
\label{Fit-inc}
\begin{tabular}{|c|c|c|c|}
\hline
                & $K_{s}^{0}$ (MeV/$c^2$) & $\phi$ (MeV/$c^2$) & $\Lambda$ (MeV/$c^2$) \\ \hline
our measurements  & $496.9 \pm0.1$  & $1019.6  \pm0.1  $  & $1115.3  \pm0.1  $ \\ \hline
PDG values      & $497.67\pm0.03$ & $1019.417\pm0.014$  & $1115.683\pm0.006$ \\ \hline
$\Delta M$      & $-0.8 \pm0.1 $ & $   0.2 \pm0.1  $  & $-0.4\pm0.1  $ \\ \hline
\end{tabular}
\end{table}

Table~\ref{Fit-error} summarizes all contributions to the systematic error of 
the mass and full-width of the $\eta_c$. 
The effect of the background shape is studied by using 
a third-order polynomial function instead of a second-order one. 
The upper fitting bound is checked by changing it from 3.05 to 3.07 GeV/$c^2$, and
the $J/\psi$ veto is removed from the event selection.
Contributions from differences of the detection efficiency
between data and Monte Carlo simulation, as well as uncertainties of the detector
mass resolution, are also listed in Table~\ref{Fit-error}.
Assuming no correlations among the above factors,  
the total systematic error on the mass and width are determined to be
1.2 MeV/$c^2$ and 7.4 MeV/$c^2$, respectively,
by a quadratic sum of all contributions.

\begin{figure}[htbp]
  \centering
  \begin{minipage}[b]{0.48\linewidth}
    \centering
    \includegraphics[width=7.2cm]{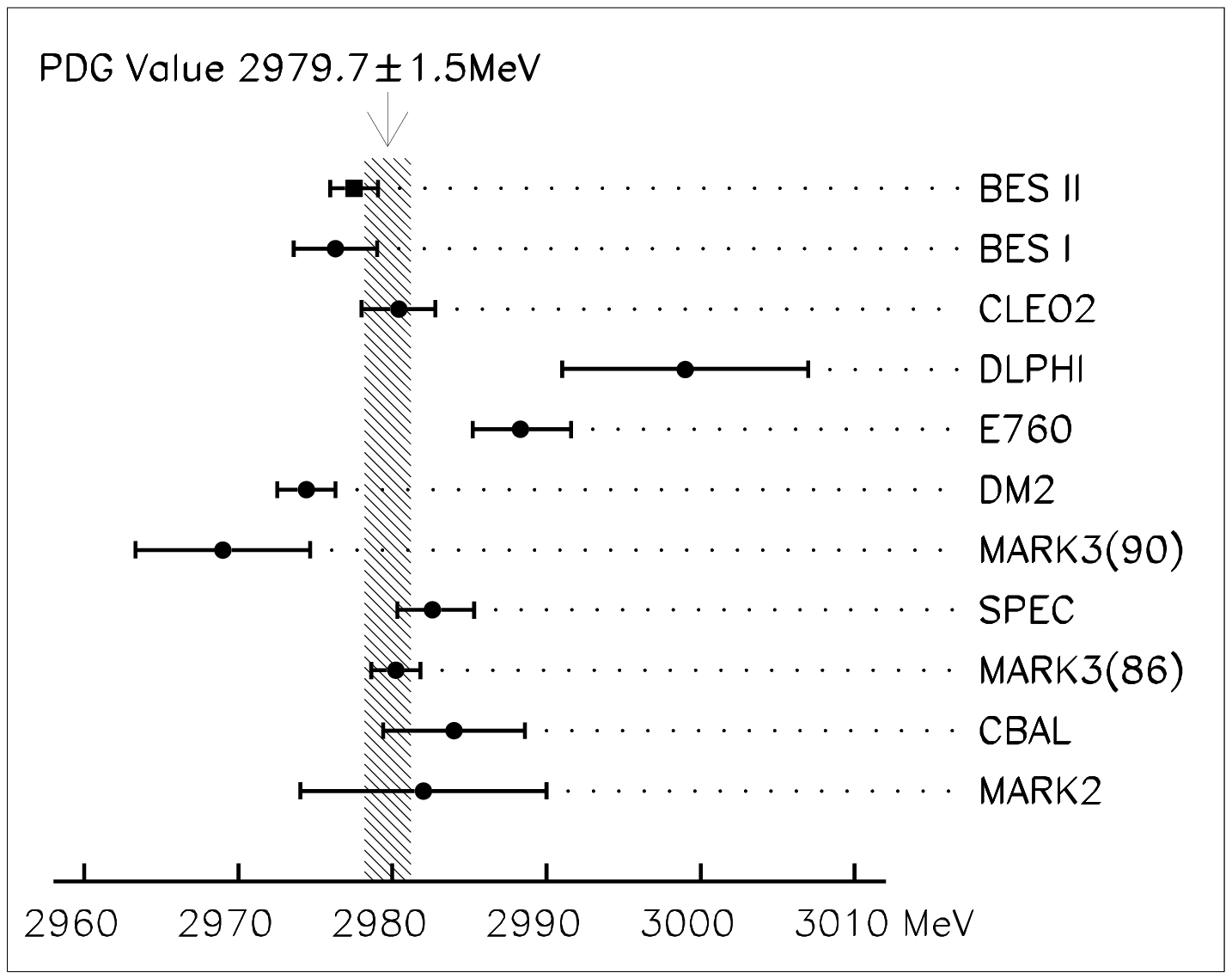}
  \end{minipage}
  \hfill
  \begin{minipage}[b]{0.48\linewidth}
    \centering
    \includegraphics[width=7.2cm]{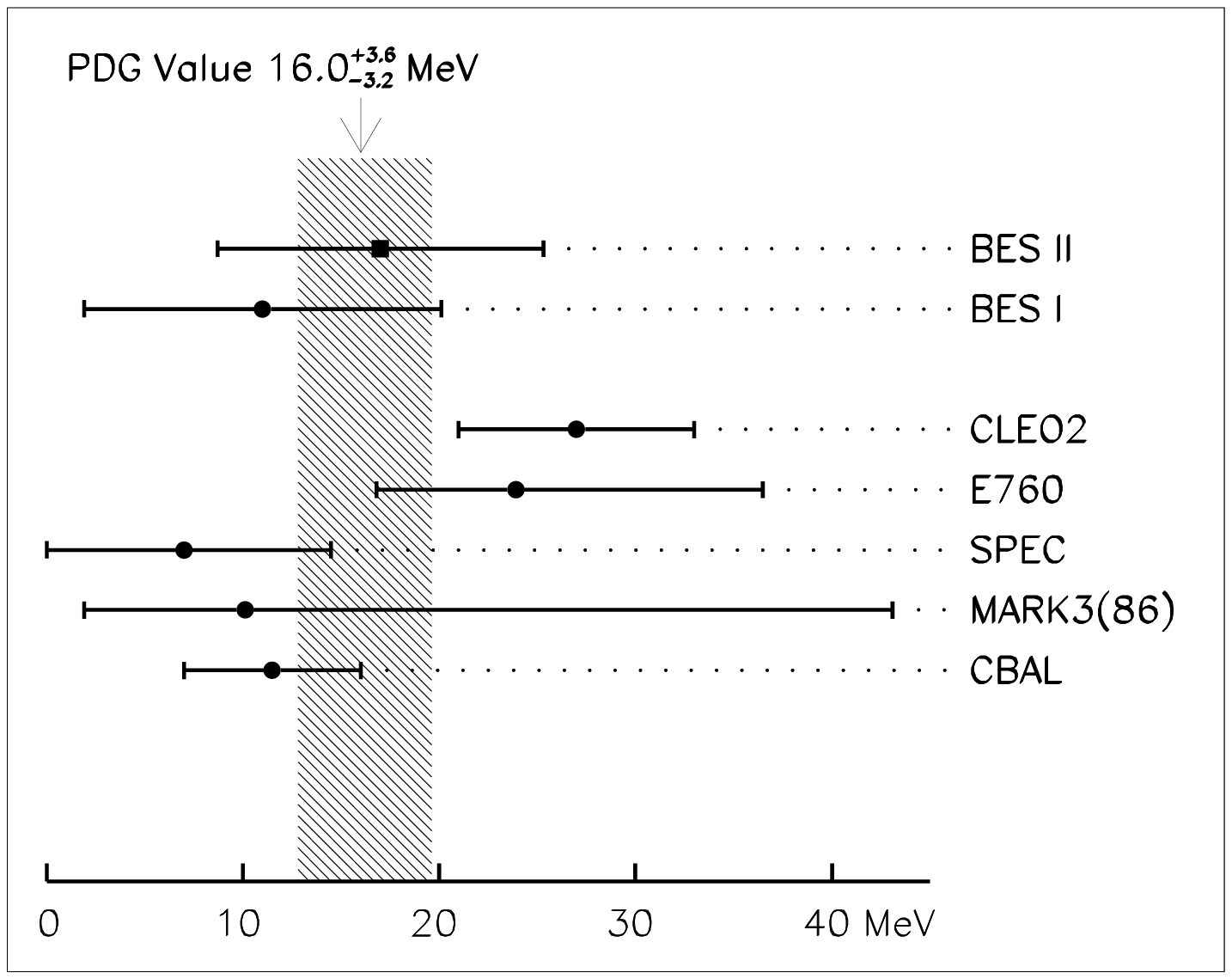}
  \end{minipage}
  \caption{Mass and full-width measurements of the $\eta_c$ meson.}
  \label{diff}
\end{figure}

In summary,
we used 58 million $J/\psi$ events collected by the BES II detector
to measure the mass and full width of the $\eta_c$ in five different
decay modes. They are determined to be
$m_{\eta_c}=2977.5\pm1.0~(\mbox{stat.})\pm1.2~(\mbox{syst.})$ MeV/$c^2$
and $\Gamma_{\eta_c} = 17.0\pm 3.7~(\mbox{stat.})\pm7.4~(\mbox{syst.})$
MeV/$c^2$. Figure~\ref{diff} shows the BES results together with previously
reported measurements.
It can be seen that the $\eta_c$ mass and width measurement from BES II
are in good agreement with the PDG averages.

\begin{table}[htb]
\caption{Sources of systematic errors.}
\label{Fit-error}
\begin{tabular}{|c|c|c|}
\hline
 sources               & error on mass & error on width \\
                       & (MeV/$c^2$)  & (MeV/$c^2$)   \\ 
\hline
mass scale calibration & 0.8           &     \\
background shape       & 0.0           & 3.6 \\ 
fitting range              & 0.3           & 2.2 \\
detection efficiency difference: data vs. MC & 0.5         & 0.3 \\
$J/\psi$ veto           &   0.7         & 5.6 \\
uncertainties of experimental mass resolution & 0.0         & 2.4 \\
\hline
total systematic error & 1.2         & 7.4 \\
\hline
\end{tabular}
\end{table}

{\vspace{0.8cm}
   The BES collaboration expresses their thanks for the hard efforts of the staff of BEPC and the 
computing center at the Institute of High Energy Physics, Beijing. 
This work is supported in part by the National Natural Science Foundation
of China under contracts Nos. 19991480,10175060
and the Chinese Academy of Sciences under contract No. KJ 95T-03(IHEP); and
by the Department of Energy under Contract Nos.
DE-FG03-93ER40788 (Colorado State University),
DE-AC03-76SF00515 (SLAC), DE-FG03-94ER40833 (U Hawaii),
DE-FG03-95ER40925 (UT Dallas).

\end{document}